\documentstyle[12pt]{article}
\newcommand{\simleq}{\; \raisebox{-0.4ex}{\tiny$\stackrel
{{\textstyle<}}{\sim}$}\;}
\newcommand{\beq}{\begin{equation}}
\newcommand{\beqar}{\begin{eqnarray}}
\newcommand{\eeq}[1]{\label{#1} \end{equation}}
\newcommand{\eeqar}[1]{\label{#1} \end{eqnarray}}

\begin{document}
\baselineskip 25pt minus .1pt
\begin{center}
{\large{\bf Isoscalar dipole strength in $^{208}_{82}$Pb$_{126}$:
the spurious mode and the strength in the continuum}} \\
[4ex]

{\bf I.Hamamoto$^{a,b,c}$ and H.Sagawa$^{d}$} \\ [2ex]

$^a$ Department of Mathematical Physics \\ Lund Institute of Technology at
University of Lund, Lund, Sweden. \\ [2ex]

$^b$ The Niels Bohr Institute\\ Blegdamsvej 17, DK-2100, Copenhagen $\O$,
Denmark. \\ [2ex]

$^c$ Radiation Laboratory, RIKEN\\ Wako-shi, Saitama 351-0198, Japan. \\  [2ex]

$^d$ Center for Mathematical Sciences \\ University of Aizu, Ikki-machi,
Aizu-Wakamatsu, Fukushima 965, Japan. \\ [2ex]

\end{center}

\vspace{2cm}

\noindent
ABSTRACT : Isoscalar dipole (compression) mode is studied first using schematic
harmonic-oscillator model and, then, the self-consistent Hartree-Fock (HF) and
random phase approximation (RPA) solved in coordinate space.
Taking $^{208}$Pb and the SkM* interaction as a numerical example, the spurious
component and the strength in the continuum are carefully examined using the sum
rules.  It is pointed out that in the continuum calculation one has to use an
extremely fine radial mesh in HF and RPA in order to separate,
with good accuracy, the spurious
component from intrinsic excitations.

\vspace{1cm}

\noindent
PACS numbers~ : ~21.10.Re, 21.60.Jz, 23.20.Js, 27.80.+w

\newpage
\section{Introduction}
\indent
Though the isoscalar (IS) dipole resonance (a compression mode)
was theoretically studied already
more than 20
years ago \cite{TJD73,HD80,GS81,DS83} and various experimental efforts
to pin down the IS giant dipole resonance (ISGDR) have been made
\cite{HPM80,CD82,GSA86,BFD97,HLC99,CLY01a},
the distribution of the IS dipole strength obtained from
the analysis of experimental data is still under dispute.
The difficulty in extracting the experimental
dipole strength comes from the ambiguity in
the parameters of the optical potentials \cite{CLY01},
which are needed in the analysis of
the relevant $(\alpha, \alpha^{'})$ scattering data.
In order to simplify our discussion, as a numerical example in the
present article we discuss
only the doubly magic nucleus $^{208}_{82}$Pb$_{126}$.
A current hot issue in both theory and experiments is whether or not a
considerable amount of
IS dipole strength is found in the energy region of $E_x$=8$-$17 MeV,
which is
much lower than the so-called ISGDR with a peak around $E_x$=23 MeV.
Available calculations have a tendency to give 1-3 MeV higher peak-energy of
ISGDR than that identified so far in experiments.

\indent
The calculation of IS dipole strength requires special care since
the spurious translation mode has to be fully
separated from intrinsic excitations.
In the self-consistent random-phase-approximation (RPA) calculation
the spurious mode should appear at zero
energy with a finite contribution to the energy-weighted sum-rule (EWSR).
The response function to the IS dipole operator
\begin{equation}
D = \sum_{k} \, (r_k^3 \, Y_{1 \mu}(\theta_k))
\label{eq:d3y1}
\end{equation}
contains the strengths of both the spurious mode and intrinsic
excitations.
In actual numerical calculations of self-consistent RPA the spurious mode
appears not at zero energy but at either a finite energy or an imaginary
energy, due to numerical inaccuracy
as well as some possible practical approximations.
This may lead to a question of whether or not
the calculated IS dipole strength
obtained in the energy region of intrinsic excitations
is free from the spurious component.
The difficulty in the present problem lies in the fact that
a very small amount of the spurious component remaining in the energy region
of intrinsic excitations
can in turn produce an appreciable amount of the strength for the operator
(\ref{eq:d3y1}) due to the power of $r^3$.
If the transition density coming from spurious component
remaining in the region is proportional to
$\frac{d \rho_0}{dr}$ where
$\rho_0$ expresses the ground-state density, the IS dipole
strength which is
free from the spurious component is obtained by using the operator
\begin{equation}
\bar{D} = \sum_{k} \, (r_k^3 - \eta \, r_k) \, Y_{10}(\theta_k)
\label{eq:b3y1}
\end{equation}
where $\eta$=$\frac{5}{3}\langle r^2 \rangle$ \cite{GS81,DS83,HSZ98},
instead of
using that in (\ref{eq:d3y1}).
However, if some spurious components remain at the energies of various
unperturbed particle-hole (ph) excitations, the relevant transition
densities are similar to those of respective original ph
excitations rather
than $\frac{d \rho_0}{dr}$.  If so, using the operator in (\ref{eq:b3y1})
may not help to take away all strength coming from the spurious component
remaining in the region of intrinsic excitations.

In ref. \cite{HSZ98} we studied both the IS and isovector (IV) dipole mode
using the
self-consistent RPA calculation with the SkM* interaction,
in which both the IS and IV correlations in RPA
are taken into account solving both the Hartree-Fock (HF)
and RPA equations in coordinate
space using the Green's function.
Examining the IS dipole strength obtained by using the operator
in (\ref{eq:b3y1}), in the energy region of $7 < Ex < 17$ MeV
we obtained an appreciable amount of IS dipole strength almost exactly at
the energies of various unperturbed ph excitations.
Since the residual interaction relevant to the IS dipole correlation
is in general attractive, the strength increased by the RPA correlation is
expected to be found at energies lower than the unperturbed ones.
Therefore, using the procedure described in ref. \cite{HSZ98}
we tried to remove the strength at finite energies, which seemed to
come from the remaining spurious component.
In the continuum at a given energy the excitations of various modes with various
quantum numbers may coexist.  The separation of the strengths coming from modes
with different quantum numbers is easy, however, we have found no established
way of separating the strengths originating from
different modes with the same quantum
numbers.
Since both the spurious mode and IS dipole resonance
are isoscalar and have the same
spin-parity, $I^{\pi} = 1^-$,
our basic idea used in ref. \cite{HSZ98} was the following.
Near the isolated resonance, $E \approx E_n$, the relation
\begin{equation}
Im \left( G_{RPA} (\vec{r}, \vec{r'}, E)  \right) \propto \rho^{tr}_{n}
(\vec{r}) \, \rho^{tr}_{n} (\vec{r'})
\end{equation}
is known.
At a high energy in the continuum, where resonances $i$ and $j$ with respective
finite widths may overlap, we may write
\begin{eqnarray}
Im \left( G_{RPA} (\vec{r}, \vec{r'}, E)  \right) & \propto &
a_i \, \rho^{tr}_{i} (\vec{r}) \, \rho^{tr}_{i} (\vec{r'}) \, + \,
b_j \, \delta \rho^{tr}_{j} (\vec{r}) \, \delta \rho^{tr}_{j}
(\vec{r'}) \nonumber\\
& \equiv & F(\vec{r},\vec{r'})
\end{eqnarray}
where the modes, $i$ and $j$,
have the same spin-parity and are diagonal in RPA.
Choosing some $f(\vec{r})$ so that
\begin{equation}
\int d\vec{r'} \, \rho^{tr}_{i} (\vec{r'}) \, f(\vec{r'}) = 0
\end{equation}
one obtains
\begin{equation}
\delta \rho^{tr}_{j} (\vec{r}) \propto
\int d\vec{r'} \, F(\vec{r},\vec{r'}) \, f(\vec{r'})
\label{eq:srho}
\end{equation}
Choosing $f(\vec{r}) = r \, Y_{1 \mu}$ in the IS dipole case, we obtained
$\delta \rho^{tr}_{j} (\vec{r})$, which
was supposed to be the transition density at the energy $E$, that carried the
spurious
$r Y_{1 \mu}$ strength.  Then, we subtracted
$\delta \rho^{tr}_{j} (\vec{r})$ from
the transition density $\rho_{tr}(\vec{r})$ at $E$
for the operator $D$ in (\ref{eq:d3y1}),
\begin{equation}
\rho_{tr}(\vec{r}) \propto \int d\vec{r'} \,
Im \left( G_{RPA}(\vec{r}, \vec{r'}, E) \right) \, (r')^3 \, Y_{1 \mu}
(\theta '_{k})
\end{equation}
Using the above subtraction procedure, it seemed that in ref. \cite{HSZ98}
we subtracted a bit too much strength as was criticized, for example, in ref.
\cite{SS02}.  The oversubtraction happened,
since the remaining ph excitations which carry some
spurious strength were not really RPA solutions.

\indent
In all available calculations of ISGDR except those
in refs. \cite{GS81,HSZ98,JP00}
the states in the continuum are approximated using the expansion
in terms of discrete basis.
Namely,
the continuum wave functions are expanded in terms of harmonic
oscillator basis with a maximum principal-quantum-number $N_{max}$.
In those discrete-basis
calculations
the operator in (\ref{eq:b3y1}) is commonly used in order to subtract the
spurious components.
The IS dipole strength obtained in the energy region of $E_x = 8-17$ MeV depends
very much on publications.  The relativistic RPA calculation in ref.
\cite{VWR00} produced especially a large amount of the IS dipole strength in the
low-energy region, for all effective interactions used.  In none of the
discrete-basis calculations the EWSR carried by the
spurious state and intrinsic excitations is carefully examined.

\indent
In the present paper we
use EWSR as a measure of the validity of numerical calculations.
The EWSR for the IS dipole operator in (\ref{eq:d3y1})
is written as
\begin{eqnarray}
S(r^3 Y_{10}) &\equiv& \sum_{n} E_n \, B(D;0 \rightarrow 1^{-} n) \nonumber\\
&=& (\frac{3}{4 \pi}) \left( \frac{\hbar ^2 A}{2M} \right)
11 \langle r^4 \rangle
\label{eq:ew3}
\end{eqnarray}
where $E_n$ expresses the energy of RPA states, while
$\langle r^4 \rangle$ on the r.h.s. is evaluated with the HF
ground state \cite{DJT61}.
The total EWSR in (\ref{eq:ew3}) consists of the contribution by
the spurious state $S_{spr}$ and that by the intrinsic excitations
$S_{intr}$, $S = S_{spr} + S_{intr}$.
The latter can be written as \cite{DS83}
\begin{eqnarray}
S_{intr}(r^3 Y_{10}) &=& \sum_{intrinsic, i} E_i \, B(D;0
\rightarrow 1^{-} i) \nonumber\\
&=& (\frac{3}{4 \pi}) \left( \frac{\hbar ^2 A}{2M} \right)
\left( 11 \langle r^4 \rangle - \frac{25}{3} \langle r^2 \rangle ^2 \right)
\label{eq:phs3}
\end{eqnarray}
which is indeed equal to the EWSR for the IS dipole operator
in (\ref{eq:b3y1}).

\indent
In sect. 2 we use a harmonic oscillator model to evaluate the distribution of
the EWSR of IS dipole strength among the spurious state,
the lower-lying and higher-lying intrinsic excitations.
This schematic model is very much
simplified, nevertheless, we expect that the semi-quantitative feature of the
result obtained from
the model should remain unchanged in realistic calculations.
In sect. 3 we present the result of the
self-consistent RPA calculation
with the Skyrme SkM* interaction on $^{208}$Pb, which is the continuum
calculation solving both HF and RPA in coordinate space.
The result is compared with discrete-basis calculations,
in which the wave functions
in the continuum are expanded in terms of
harmonic oscillator basis.
In sect.4 conclusion and discussions are given.

\section{Harmonic oscillator model}
\indent
The distribution of IS dipole strength may be studied in terms of
a simplified model \cite{BM75}
in which the particle motion is described by a harmonic
oscillator potential.
The particle excitations produced by the field (\ref{eq:d3y1})
are governed by the selection
rule $\Delta N$=1 or 3 and have the energies $\hbar \omega_0$ and
$3 \hbar \omega_0$, respectively.
For a single completed shell with principal quantum number $N$, the transition
strength is given by
\begin{equation}
\sum_{\nu_N, \nu_{N+1}} |\langle \nu_{N+1} | r^3 Y_{10} | \nu_N \rangle |^2 =
\frac{3}{4 \pi}
\left(\frac{\hbar}{2M \omega_0}\right)^3 \frac{1}{12} (N+1) (N+2) (N+3)
(13 N^2 + 52 N +50)
\label{eq:n1}
\end{equation}
\begin{equation}
\sum_{\nu_N, \nu_{N+3}} |\langle \nu_{N+3} | r^3 Y_{10} | \nu_N \rangle |^2 =
\frac{3}{4 \pi} \left(\frac{\hbar}{2M \omega_0} \right)^3 \frac{1}{12}
(N+1) (N+2) (N+3) (N+4) (N+5)
\label{eq:n3}
\end{equation}
where $\nu_N$ represents the quantum numbers needed to specify the
single-particle states in the shell $N$.  For simplicity, we do not include
the spin-isospin degeneracy in the present model.
Since the last three occupied
shells contribute to the total $\Delta N$=3 strength we
obtain
\begin{equation}
\sum_{N=N_F-2}^{N_F} \; \sum_{\nu_N, \nu_{N+3}} |\langle \nu_{N+3} |
r^3 Y_{10} | \nu_N \rangle |^2
= \frac{3}{4 \pi} \left(\frac{\hbar}{2M \omega_0} \right)^3 \frac{1}{12}
(N_F +1) (N_F +2) (N_F +3) (3N_{F}^{2} + 12N_F +20)
\label{eq:tn3}
\end{equation}
while only the last-filled $N=N_F$ shell contributes
to the $\Delta N$=1 strength.

Now, the major part of the $\Delta N$=1 strength, which is obtained by setting
$N=N_{F}$ in (\ref{eq:n1}),
corresponds to the spurious
excitation, namely center of mass motion, while in the present harmonic
oscillator model the $\Delta N$=3 strength contains no spurious component.
Subtracting the strength coming from the spurious excitation
\begin{equation}
\frac{3}{4 \pi} \left(\frac{\hbar}{2M \omega_0} \right)^3 \frac{25}{24}
(N_F +1) (N_F +2)^3 (N_F +3) ,
\label{eq:cmtr}
\end{equation}
from the expression (\ref{eq:n1}) with $N=N_F$, we obtain
the strength of the intrinsic $\Delta N$=1 excitation
\begin{equation}
\frac{3}{4 \pi} \left(\frac{\hbar}{2M \omega_0} \right)^3 \frac{1}{24}
N_F (N_F +1) (N_F +2) (N_F +3) (N_F +4)
\label{eq:tn1}
\end{equation}
We note that the total number of particles in the present model is written as
\begin{equation}
\frac{1}{6} (N_F +1) (N_F +2) (N_F +3)
\label{eq:pn}
\end{equation}
while
\begin{equation}
\sum_{N=0}^{N_F} \langle r^2 \rangle
= \left(\frac{\hbar}{2M \omega_0} \right) \frac{1}{4} (N_F +1) (N_F +2)^2
(N_F +3)
\label{eq:r2}
\end{equation}
and
\begin{equation}
\sum_{N=0}^{N_F} \langle r^4 \rangle
= \left(\frac{\hbar}{2M \omega_0} \right)^2 \frac{1}{2} (N_F +1) (N_F +2)
(N_F +3) (N_F^2 + 4N_F +5)
\label{eq:r4}
\end{equation}

\indent
The spurious contribution (\ref{eq:cmtr}) is estimated
by recognizing that the spurious component is found only at the $\Delta N$=1
excitations and that the transition density of the spurious excitation is
proportional to the
radial derivative of the ground-state density, $d \rho_0 / dr$.

\indent
From the expressions, (\ref{eq:tn3}) and (\ref{eq:tn1}),
one obtains the ratio of the unperturbed transition strength of $\Delta N$=1
excitations to that of  $\Delta N$=3 ones to be 1:6 for $N_F \gg 1$.
Correspondingly,
the ratio of the unperturbed EWSR is 1:18.

\indent
Introducing the residual interaction of a separable form
\begin{equation}
V = c \, D^{\dagger} D
\label{eq:resv}
\end{equation}
where $c$ expresses an attractive coupling constant,
we solve the RPA equation and
estimate the distribution of
the transition strength together with the RPA energies.
The following results of cases (a) and (b) are summarized in Table I. \\
(a) If we choose $c$ so that the higher-lying RPA solution is obtained at
(2.5)$\hbar
\omega_0$, the lower-lying RPA solution is found at (0.872)$\hbar \omega_0$.
Then, the RPA transition strength is distributed with 1.09:2.79
over the lower-lying and higher-lying RPA solutions,
while the distribution of the EWSR is 0.95:6.98 . \\
(b) If we choose $c$ so that the higher-lying RPA solution is obtained at
(2.8)$\hbar
\omega_0$, the lower-lying RPA solution is found at (0.961)$\hbar \omega_0$.
Then, the RPA transition strength is distributed with 0.60:2.62
between the lower-lying and higher-lying RPA solutions,
while the distribution of the EWSR is 0.58:7.34 .

\indent
Comparing the peak energies of the RPA solutions
with those obtained by available self-consistent HF
plus RPA calculations, one expects that the realistic situation may lie
somewhere between the case (a) and (b).
Before the spurious component $\delta \, \rho _{j}^{tr} (\vec{r})$ in
(\ref{eq:srho}) was subtracted
in our self-consistent RPA calculations
of ref. \cite{HSZ98},
a number of
lower-energy peaks at Ex$\simleq$17 MeV for the operator
(\ref{eq:b3y1}) were found nearly at respective unperturbed ph energies.
In contrast, the higher-energy peak around 25 MeV
was collectively constructed by shifting the unperturbed
strength coherently from the region of higher energy.
In other available self-consistent RPA calculations, for example
in refs. \cite{co00} and
\cite{VWR00} in which the expansion in terms of harmonic oscillator basis
was used,
or in ref. \cite{JP00} which was a continuum calculation of relativistic
Hartree plus RPA calculation, a considerable amount of
low-energy strength appeared also around unperturbed ph energies.
Compared with those HF plus RPA calculations,
in the case (a) of Table I the higher-lying RPA peak
is pushed down by about the same order of magnitude and
the lower-lying peak is too strongly pushed down, while
in (b) the higher-lying peak is appreciably less pushed down and
the lower-lying peak is still slightly lower.

On the other hand, one might wonder whether or not the ratio of
the coupling between the $\Delta N$=1 and 3 excitations
to that within the excitations of a given $\Delta N$ is realistic in the
present schematic interaction of (\ref{eq:resv}).
In the present harmonic-oscillator model
the effective strength of the coupling
between the $\Delta N$=1 and 3 excitations can be measured from
the increase of EWSR in the lower-lying RPA solution compared with
that in the unperturbed $\Delta N$=1 excitation.
The portion of the EWSR in (\ref{eq:ew3}) carried by the lower-energy peak
is increased from 2.3 to 5.2 percent in the case (a)
and to 3.2 percent in the case (b), as shown in Table I.

\section{Hartree-Fock and RPA calculations with Skyrme interactions}

In the HF calculation with the SkM* interaction for $^{208}$Pb we obtain
\begin{eqnarray}
\langle r^2 \rangle &=& 30.85 \quad fm^2 \nonumber\\
\langle r^4 \rangle &=& 1218.2 \quad fm^4
\label{eq:r2r4}
\end{eqnarray}
Thus, using eqs. (\ref{eq:ew3}) and (\ref{eq:phs3}) the EWSR becomes
\begin{equation}
S(r^3 Y_{10}) \; = \; 13.79 \, \times \, 10^6  \quad fm^6 \, MeV
\label{eq:ewn}
\end{equation}
\begin{equation}
S_{intr}(r^3 Y_{10}) \; = \; 5.632 \, \times \, 10^6  \quad fm^6 \, MeV
\label{eq:ewn1}
\end{equation}
From eqs. (\ref{eq:ewn}) and (\ref{eq:ewn1})
the contribution by
the spurious state to the EWSR for the operator $r^3 Y_{1 \mu}$ is
8.16 $\times$ 10$^6$ $fm^6 \, MeV$, which is 59.2 percent of the total EWSR in
(\ref{eq:ewn}).

In the present work we perform the self-consistent RPA calculation
with the SkM* interaction, in which both the IS and IV correlations in RPA are
taken into account solving both HF and RPA equations in coordinate space using
the Green's function.  The difference of the present calculation from that
in ref. \cite{HSZ98} is that we use the radial mesh ($\Delta r$) of 0.1 $fm$
in both the HF and
RPA calculations and, moreover, we carefully check the sum-rules consumed by
both intrinsic and spurious excitations.  We present the numerical result without
adopting the subtraction
procedure, which was used in ref. \cite{HSZ98} and was also explained in the
Introduction.
In ref. \cite{HSZ98} $\Delta r$=0.1 $fm$ was used in HF, while we adopted
$\Delta r$=0.3 $fm$ in RPA.
In Fig. 1 we show the RPA response functions, which are smeared out using the
width of 0.5 MeV, for the operator
(\ref{eq:d3y1}) obtained with $\Delta r$=0.1 $fm$ and 0.3 $fm$ and for the
operator (\ref{eq:b3y1}) calculated with $\Delta r$=0.1 $fm$.
A great, somewhat unexpected improvement in the result of
the present calculation with a finer
radial mesh in RPA is :
the appreciable difference between the response functions to the operators
(\ref{eq:d3y1}) and (\ref{eq:b3y1}) in the energy region of
the higher peak around
25 MeV is drastically decreased.
Moreover, the spurious state appears now slightly below 0 MeV (namely, at an
imaginary energy)
and consumes a much larger portion
of EWSR, $S_{spr}$.

As in ref. \cite{HSZ98},
our effective interaction used in RPA consists only of the part in
the spin-independent channels.
Using discrete-basis calculations, in which the inclusion of
the full Skyrme effective
interaction is straightforward and easy, we have checked
that the RPA response function of the present IS dipole operator
changes very little when we use the full effective interaction,
as seen in Tables II and III.

In Table II we show the transition strength for the operator (\ref{eq:d3y1}),
the contribution to EWSR,
and the percent of the total EWSR, $S(r^3Y_{10})$, at the spurious state and
in the region of $2 < E_x < 17$ MeV, $17 < E_x < 70$ MeV, and
$E_x > 70$ MeV.
In Table III the same quantities are tabulated for the operator (\ref{eq:b3y1}).
The borderline 17 MeV is chosen, since almost all $\Delta N =1,
I^{\pi} = 1^{-}$ ph excitations in the HF potential lie below 17 MeV and no
unperturbed 1$^-$ ph excitations are found for $18 < E_x < 24$ MeV.

As is previously known, it is seen from Table II that the portion of the EWSR
found in the energy region of intrinsic excitations ($E_x > 2$ MeV)
clearly shows the presence of the spurious components remaining in the region.
From Table III it is seen that the ratio of the EWSR lying in the region of
$E_x < 17$ MeV to that for $E_x > 17$ MeV is 1:4.9.
More importantly, the portion of the EWSR obtained for $E_x > 2$ MeV is
39.1 percent of the total EWSR, compared with the full value, 40.8 percent.
It should be mentioned that in the continuum calculation it is technically very
difficult to obtain accurate values of the transition strength carried by
very sharp peaks including the spurious state, in contrast to the
discrete-basis calculations.
Furthermore, in the present continuum calculation the spurious state appears
slightly below 0 MeV.  Thus, reducing the effective
ph interactions by multiplying a
factor of 0.9902, we obtain the spurious state at $E_x$=0.107 MeV, from
which we estimate the carried transition strength.
Though it is clear that the continuum calculations performed using a much finer radial
mesh than 0.1 $fm$ would further improve the accuracy,
here we do not pursue it any further.

In Fig. 2 we show the calculated RPA response functions to the operators,
(\ref{eq:d3y1}) and (\ref{eq:b3y1}), as a function of excitation energy.
Though the peaks at $E_x >$ 7.5 MeV lie in the continuum, calculated
lower-lying peaks are too sharp to be plotted.
Since the spreading width is anyhow not included in the RPA response function,
in Fig. 2 we plot the quantities, which are obtained by smearing out the
calculated RPA response function with the width of 1 MeV.
It is seen that an appreciable amount of
IS dipole strength is present in the energy
region of $E_x$=7-17 MeV, which is much lower than the so-called ISGDR around
25 MeV.  The presence of the lower-lying IS dipole strength indicates that the
ISGDR appearing around 25 MeV
is not really a good collective mode in the nuclear system.

For reference, in Tables II and III we show
also the result of discrete-basis calculations for the operators,
(\ref{eq:d3y1}) and (\ref{eq:b3y1}).
How to truncate the space of harmonic oscillator basis seems to depend on
publications.  What we have adopted in the present work, which is
along the line of ref. \cite{co00}, is the following.
First we construct the HF potential, which is calculated in coordinate space
and, then,
calculate the one-particle energies and wave-functions for the potential
using the harmonic oscillator basis with the radial node
$n^{HO} \leq n_{max}^{HO}$, where $n_{max}^{HO}$ is taken to be independent
of $\ell$.
Then, the RPA equation is solved including the lowest-lying
$\frac{1}{2} (N_{max}^{HF} - \ell)$ one-particle levels for a given $\ell$,
where $N_{max}^{HF}$
is taken to be independent of $\ell$.
This truncation of the ph space included in RPA
is approximately the same as that
limiting the ph space by the $\ell$-independent maximum energy of
particle configurations.
It is noted that, as $n_{max}^{HO}$ increases, the bound
one-particle energies and wave-functions approach the eigen energies and
eigen functions of the HF potential, respectively,
while the properties of the positive-energy one-particle levels
can be far away from the correct ones unless $n_{max}^{HO}$ is taken to be
$\infty$.  In actual calculations one must take finite values of
$n_{max}^{HO}$ and $N_{max}^{HF}$ and, thus, it is not guaranteed that using
finite larger values of $n_{max}^{HO}$ and $N_{max}^{HF}$ produces a more
accurate result of IS dipole strength.

For the parameter set, $n_{max}^{HO} = 12$ and $N_{max}^{HF} = 24$,
in Tables II and III we show both
the result obtained by using the full effective interaction
in RPA and the one by employing the same effective interaction used in our
continuum RPA calculation.
The fact that very little difference is found between these two results
indicates that the IS dipole strength obtained from
our continuum RPA calculation is reliable.
Comparing the numbers in the last column of Table II with those
in Table III, in the discrete-basis calculation
the difference between the strengths
calculated with the operators (\ref{eq:d3y1}) and (\ref{eq:b3y1})
is relatively minor in the region of $E_x > 17$ MeV.
On the other hand,
from Table II we notice that the calculated energy of the spurious state
is pretty high compared with the continuum calculation and, furthermore,
only 92 percent of
the correct portion of EWSR is consumed by the calculated spurious state.
The properties of the spurious state converge surprisingly slowly to the right
ones, as we increase $n_{max}^{HO}$ and $N_{max}^{HF}$.  On the other hand,
if we use a value of $n_{max}^{HO}$ larger than 12,
the calculated response function in the continuum starts to behave strangely
instead of converging to some reasonable result.
This strange
behavior of discrete-basis calculations seems to be known for experts in the
fields \cite{GC02}.

\section{Conclusion and discussions}

First, the distribution of IS dipole strength for heavy nuclei is studied using
the schematic harmonic-oscillator model, in which the strength is distributed
among the spurious mode, the low-energy and high-energy peaks.
Though 96 percent of the unperturbed $\Delta N=1$ strength
for the operator $r^3 Y_{1 \mu}$ corresponds to the spurious
excitation, the remaining 4 percent contributes to the intrinsic excitations.
The spurious state consumes 56.8 percent of the total EWSR for the operator
$r^3 Y_{1 \mu}$, while the low-energy RPA peak may carry up till 5 percent
of the EWSR.

Secondly, the result of IS dipole strength in $^{208}$Pb is presented, which is
calculated using the self-consistent HF plus RPA solved in
coordinate space using the Green's function method, with better accuracy than in
ref. \cite{HSZ98}.
If one performs numerical calculations with very good accuracy,
the IS dipole strength of intrinsic excitations should not
depend on whether the operator (\ref{eq:d3y1}) or (\ref{eq:b3y1}) is used.
At the same time,
the spurious mode should appear close to zero energy and consume the whole
portion of EWSR expected.
In our present continuum calculation we could not attain such a satisfactory
accuracy.
On the other hand, if the calculated strengths with the operators
(\ref{eq:d3y1}) and (\ref{eq:b3y1}) are appreciably different from each other in
the energy region of intrinsic excitations, the spurious component remaining in
the region may not have the transition density proportional to
$\frac{d \rho_{0}(r)}{dr}$.
In the present work we have not attempted to take away
this kind of possible
spurious component from the region of intrinsic excitations, since we find no
reliable way of excluding it.

Using the radial mesh of 0.1 $fm$ in both the HF and RPA
calculations,
the spurious state
is obtained slightly below 0 MeV.  Multiplying the effective ph interaction by
a factor of
0.9902, we push up the energy to 0.107 MeV and estimate the EWSR carried by the
state, which
turns out to be the major portion of the expected EWSR.
On the other hand,
the EWSR carried by the calculated states for
$E_x > 2$ MeV is 114 and 96 percent of the expected EWSR for the operators
(\ref{eq:d3y1}) and (\ref{eq:b3y1}), respectively.
The accuracy of the continuum calculations is appreciably improved by decreasing
$\Delta r$ used in solving RPA from 0.3 $fm$ in ref. \cite{HSZ98} to 0.1 $fm$,
however, it is still away from being very satisfactory.
In the continuum calculations
the accuracy of separating the spurious component from
intrinsic excitations is surprisingly sensitive to the size of radial mesh
employed in RPA.  Thus, the response function of IS dipole strength
has to be
calculated using a very fine radial mesh, in order to obtain a reliable
numerical result.
From the present study we find that slightly more than 5 percent of the total
EWSR is expected as intrinsic excitations lying in the low-energy region below
$E_x < 17$ MeV.

The expansion of HF and RPA solutions in terms of harmonic
oscillator basis can work for bound states if one includes a sufficiently large
number of bases \cite{BG77}.
Thus, in the present study of IS dipole strength
the properties of the
spurious mode become better and better as the number of bases increases,
though we find that the convergence of the properties is unexpectedly slow as
$n_{max}^{HO}$ and $N_{max}^{HF}$ increase.
On the portion of EWSR found in the continuum the discrete-basis calculation
presented in Table II and Fig. 1 produces 9 percent larger in the
$2 < E_x < 17$ MeV region and 6 percent smaller for $E_x > 17$ MeV,
compared with our
continuum calculation.   However, the real problem is :
If we further enlarge the space of discrete-basis, the calculation
starts to produce
more than 100 percent of EWSR, which are certainly unreasonable.
Thus, it is not clear whether or not the discrete-basis calculations can be
reliably used for the estimate of IS dipole strength.

\indent
The authors are grateful to Drs. N. Van Giai and G. Colo for the generous advice
on the discrete-basis calculation.
I.H. acknowledges the financial supports provided
by Crafoordska Stiftelsen and RIKEN.
H.S. acknowledges financial support by the Japanese Ministry of Education,
Science, Sports and Culture by Grant-In-Aid for Scientific Research under the
program number C(2) 12640284.

\newpage

\noindent
{\bf\large Figure captions}\\

\begin{description}
\item[{\rm Figure 1 :}]
RPA response functions to the IS dipole operators,
(\ref{eq:d3y1}) and (\ref{eq:b3y1}),
as a function of excitation energy, which are
obtained from the self-consistent HF plus RPA calculations solved in coordinate
space.  A radial mesh $\Delta r$=0.1 $fm$ is used in HF, while the results
obtained by using $\Delta r$=0.1 and 0.3 $fm$ in RPA are compared.
We show the calculated response functions, which are smeared out
using the width of 0.5 MeV.

\end{description}

\begin{description}
\item[{\rm Figure 2 :}]
RPA response functions to the IS dipole operators,
(\ref{eq:d3y1}) and (\ref{eq:b3y1}),
as a function of excitation energy, which are
obtained from the self-consistent continuum calculation (CRPA)
with $\Delta r$=0.1 $fm$ and the
discrete-basis calculation with $n_{max}^{HO} =12$ and $N_{max}^{HF} =24$
(DRPA).  In all calculations
the (1, $\tau \cdot \tau$) effective interaction (see the caption to Table II)
is used.
In order to see more easily the total distribution of the strength,
we plot both CRPA and DRPA response functions, which are
smeared out using the width of 1 MeV.

\end{description}

\newpage
\begin{table}[hbt]
\caption{Energy, reduced transition strength $B(D)$, and contribution to the
EWSR for the operator $r^3 Y_{1 \mu}$, which are calculated
in the harmonic oscillator model for
$N_F \gg 1$.  B(D) and EWSR are given in arbitrary unit.}
\vspace*{2pt}

\begin{tabular}{c|c|c|c}  \hline
 Case & energy in $\hbar \omega_0$  &   B(D)
  &  EWSR ($\%$)  \\
\hline
  unperturbed & spurious (1.0) & $\frac{25}{24} = 1.042$ & $\frac{25}{24} =
  1.042$ (56.8) \\
        & 1.0 & $\frac{1}{24} = 0.042$ & $\frac{1}{24} = 0.042$ (2.3)\\
        & 3.0   & $\frac{1}{4} = 0.25$ & $\frac{3}{4} = 0.75$ (40.9)\\
   & & & \\
  (a) & 0.0 & $\infty$ &  1.042 (56.8) \\
        & 0.872 & 0.109 & 0.095 (5.2)\\
        & 2.5   & 0.279 & 0.698 (38.0)\\
   & & & \\
  (b) & 0.0 & $\infty$ &  1.042 (56.8) \\
        & 0.961 & 0.060 & 0.058 (3.2)\\
        & 2.8   & 0.262 & 0.734 (40.0)\\  \hline

\end{tabular}

\end{table}

\begin{table}[hbt]
\caption{Energy, reduced transition strength $B(D)$, and contribution to the
EWSR for the operator $r^3 Y_{1 \mu}$ in
$^{208}$Pb, which are calculated in the self-consistent RPA
using the SkM* interaction. Inside the bracket of the forth column
the ratio to the total EWSR for the operator $r^3 Y_{1 \mu}$, (\ref{eq:ewn}),
is expressed in percent.
For discrete calculations with $n_{max}^{HO} =12$ and $N_{max}^{HF} =24$
we show two sets of
calculated results,
the one obtained by using the full effective interaction in RPA
and the other calculated by using only the spin-independent part as that
employed in our continuum RPA calculation.  The former is denoted by ``full'',
while the latter is marked by ``(1, $\tau
\cdot \tau$)'' .  In the continuum calculation it is technically very difficult
to estimate an accurate value of the transition strength carried by the spurious
state, which is given with ``$\approx$'' in the third and fourth column.  See
the text for details.}
\vspace*{2pt}

\begin{tabular}{c|c|c|c}  \hline\hline
  Type of calculation & energy in MeV  &   B(D) in $fm^6$
  &  EWSR in $fm^6 MeV$ ($\%$)  \\
\hline\hline
  correct calculation & 0.0 &$\infty $ & 8.16$\times$ 10$^6$ (59.2) \\
        & $0<E_x$ & & 5.63$\times$  10$^6$ (40.8)\\
      \hline
  continuum & 0.107 & $\approx$ 7.5$\times$  10$^7$ & $\approx$ 8.0$\times$
  10$^6$  \\
  (1, $\tau \cdot \tau $)  &  $2<E_x<17$   &   1.275  $\times$  10$^5$  &    1.433$\times$10$^6$ (10.4)           \\
        &  $17<E_x<70$  & 1.973  $\times$  10$^5$   & 4.856$\times$  10$^6$ (35.2)  \\
        &  $70<E_x<150$  &  1.3  $\times$   10$^3$  &  1.14 $\times$   10$^5$ (0.83)  \\\hline
  discrete ($n^{HO}_{max}=12, N^{HF}_{max} = 16$) & 1.401 &  5.342 $\times$   10$^6$  & 7.485 $\times$  10$^6$ (54.7) \\
 full
           &  $2<E_x<17$    &  1.347 $\times$   10$^5$ & 1.534  $\times$   10$^6$ (11.1)   \\
        &  $17<E_x<70$   &  1.911 $\times$   10$^5$  & 4.677  $\times$   10$^6$(33.9)   \\
       &  $70<E_x<88$   &  52.8   & 3.869 $\times$   10$^3$(0.028)   \\ \hline
  discrete ($n^{HO}_{max}=12, N^{HF}_{max} = 24$) & 0.758 &  9.949 $\times$   10$^6$  & 7.541 $\times$  10$^6$ (54.7) \\
  full
           &  $2<E_x<17$    &  1.346 $\times$   10$^5$ & 1.527  $\times$   10$^6$ (11.1)   \\
        &  $17<E_x<70$   &  1.914 $\times$   10$^5$  & 4.683  $\times$   10$^6$(34.0)   \\
       &  $70<E_x<190$  &  476.8   & 4.141  $\times$   10$^4$(0.30)     \\\hline
  discrete ($n^{HO}_{max}=12, N^{HF}_{max} = 24$) & 0.406 & 1.850 $\times$   10$^7$ & 7.538 $\times$   10$^6$ (54.7)  \\
 (1, $\tau \cdot \tau $)       &  $2<E_x<17$   & 1.411 $\times$   10$^5$ & 1.560  $\times$   10$^6$ (11.3) \\
        &  $17<E_x<70$   & 1.899  $\times$   10$^5$  & 4.653$\times$   10$^6$ (33.7) \\
   & $70<E_x<190$   &  475.6  & 4.140 $\times$   10$^4$(0.30)
     \\ \hline \hline

\end{tabular}

\end{table}

\begin{table}[hbt]
\caption{The same quantities as in Table II, but
for the operator $(r^3 - \eta \, r) Y_{1 \mu}$ in
$^{208}$Pb.  See the caption to Table II.}
\vspace*{2pt}

\begin{tabular}{c|c|c|c}  \hline\hline
  &   &   & \\
  Type of calculation & energy in MeV  &   B($\bar{D}$) in $fm^6$
  &  EWSR in $fm^6 MeV$ ($\%$)  \\
\hline\hline
  correct calculation & 0.0 & 0.0 & 0.0 (0) \\
        & $0<E_x$ & & 5.63$\times$   10$^6$ (40.8)\\\hline
  continuum & 0.107 & 1.163 $\times$   10$^4$ & 1.240 $\times$   10$^3$ (0.009)\\
  (1, $\tau \cdot \tau $)  &  $2<E_x<17$   & 8.294  $\times$   10$^4$  &  9.064 $\times$   10$^5$ (6.6) \\
        &  $17<E_x<70$  &  1.802 $\times$  10$^5$  & 4.440 $\times$  10$^6$ (32.2)  \\
        &  $70<E_x<150$  &  0.005 $\times$ 10$^5$  &  0.042 $\times$ 10$^6$ (0.3)  \\\hline
 discrete ($n^{HO}_{max}=12,N^{HF}_{max} = 16$) & 1.401 & 1.672 $\times$   10$^3$   & 2.342 $\times$   10$^3$ (0.011)  \\
  full        &  $2<E_x<17$   &  9.369 $\times$   10$^4$ &  1.006 $\times$   10$^6$  (7.3)  \\
        &  $17<E_x<70$   &  1.878 $\times$   10$^5$  &  4.580 $\times$   10$^6$  (33.2) \\
       &  $70<E_x<88$   &  36.2   &  2.653 $\times$   10$^3$  (0.019) \\ \hline
 discrete ($n^{HO}_{max}=12,N^{HF}_{max} = 24$) & 0.758 & 2.005 $\times$   10$^3$   & 1.520 $\times$   10$^3$ (0.011)  \\
  full        &  $2<E_x<17$   &  9.492 $\times$   10$^4$ &  1.015 $\times$   10$^6$  (7.4)  \\
        &  $17<E_x<70$   &  1.889 $\times$   10$^5$  &  4.604 $\times$   10$^6$  (33.4) \\
       &  $70<E_x<190$   &  368.1  &  3.165 $\times$   10$^4$  (0.23) \\ \hline
 discrete ($n^{HO}_{max}=12,N^{HF}_{max} = 24$) & 0.406 & 3.607 $\times$   10$^3$   & 1.464 $\times$   10$^3$ (0.011)  \\
  (1, $\tau \cdot \tau $)        &  $2<E_x<17$   &  9.544 $\times$   10$^4$ &
  1.023 $\times$   10$^6$  (7.4)  \\
        &  $17<E_x<70$   &  1.886 $\times$   10$^5$  &  4.597 $\times$   10$^6$  (33.3) \\
       &  $70<E_x<190$   &  364.7 &  3.130 $\times$   10$^4$  (0.23) \\ \hline\hline
\end{tabular}

\end{table}

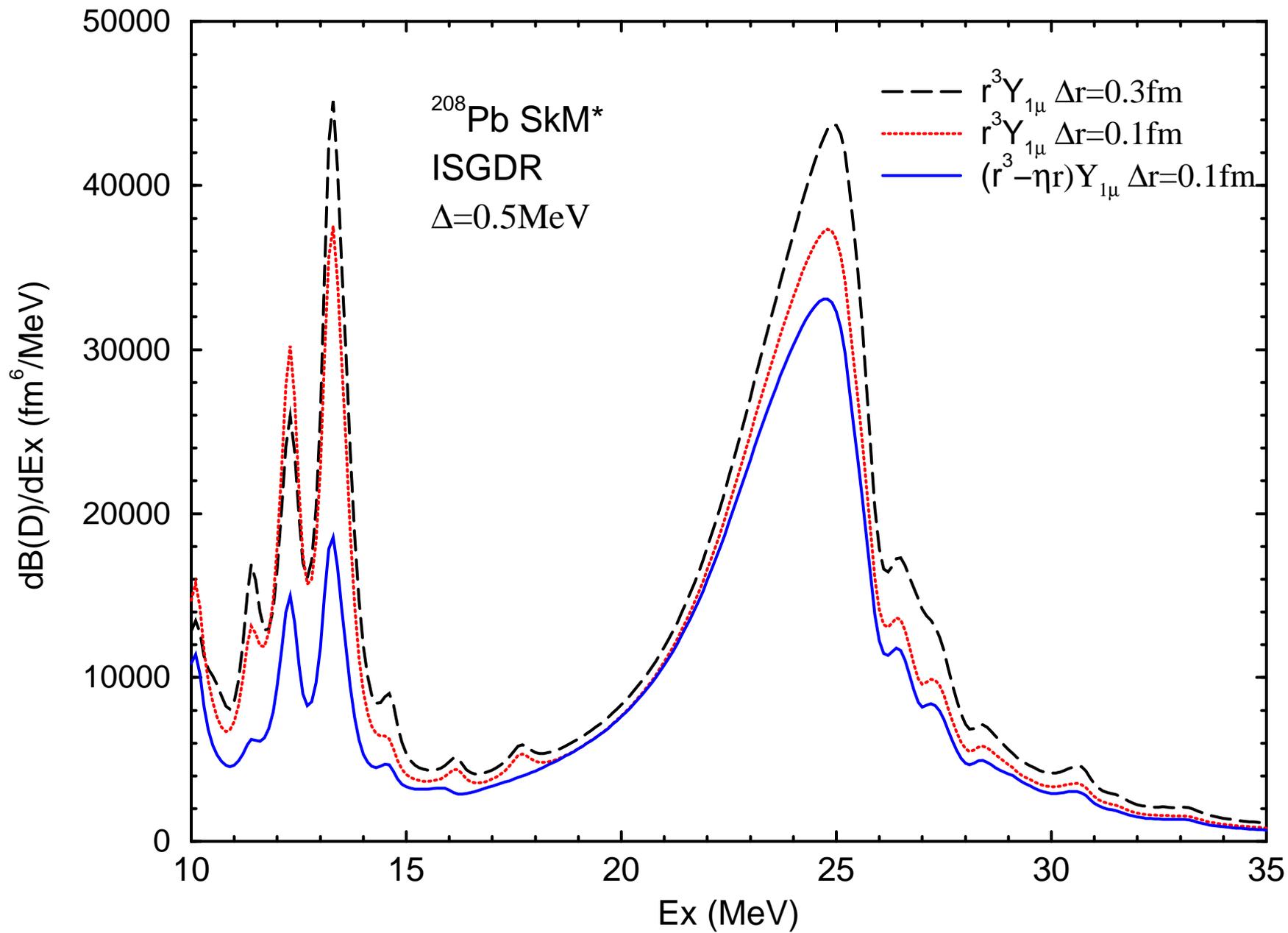
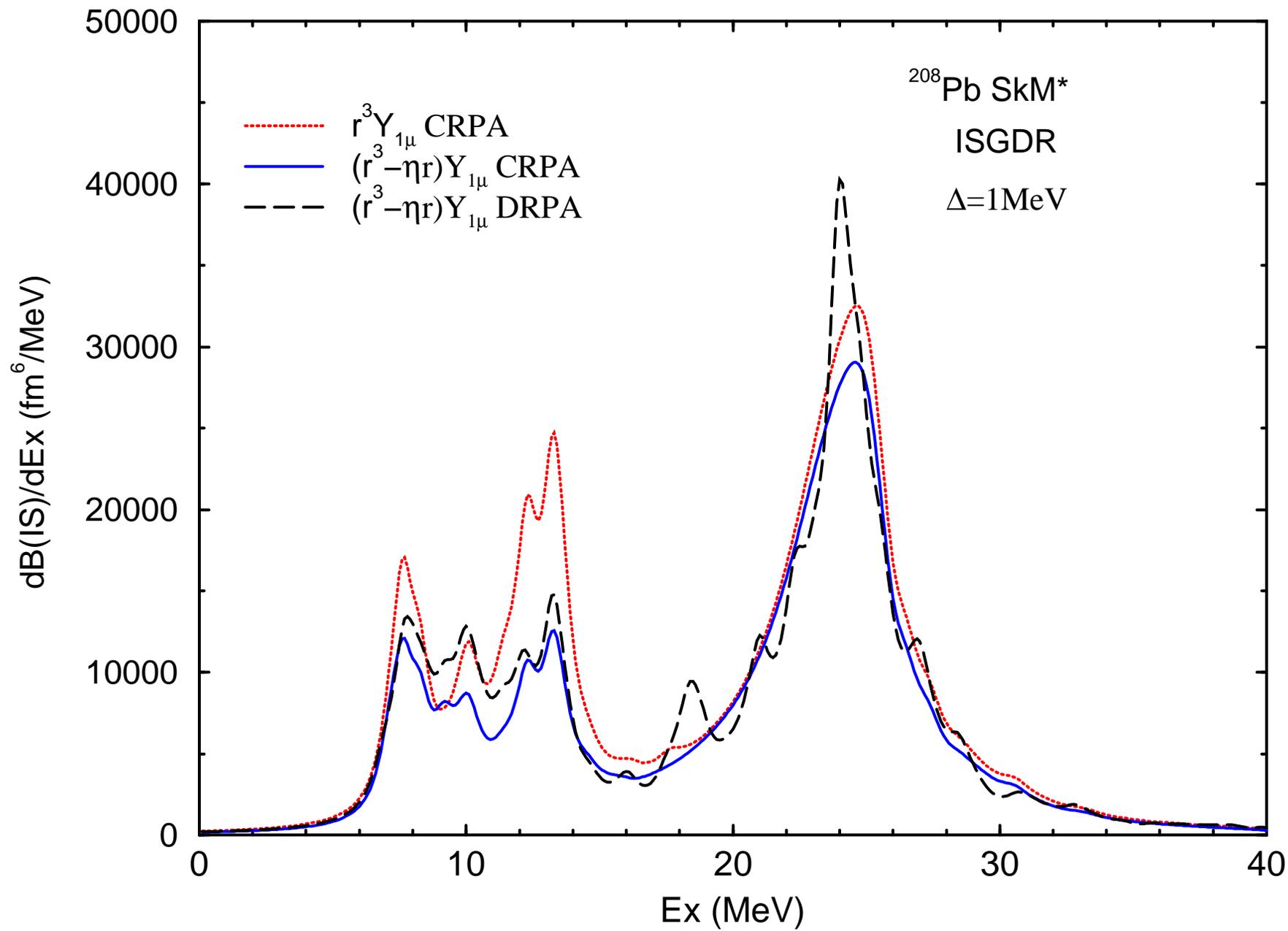
\end{document}